\documentstyle[12pt]{article}
 
\begin{document} 

\baselineskip 22pt 

\begin{center}
{\Large 
\bf Decay Constants of $B$, $B^*$ and $D$, $D^*$ Mesons\\ 
in Relativistic Mock Meson Model}\\
\vspace{1.0cm}
Dae Sung Hwang$^1$ and Gwang-Hee Kim$^2$\\
{\it{Department of Physics, Sejong University, Seoul 143--747,
Korea}}\\
\vspace{2.0cm}
{\bf Abstract}\\
\end{center}

We derive formulas for the decay constants $f_P$ and $f_V$
of pseudoscalar and vector mesons in the relativistic
mock meson model.
Using these formulas, we obtain $f_P$ and $f_V$ of
$B_s$, $B_d$, $D_s$, and $D_d$ mesons as functions of
the mock meson parameter $\beta$.
Then by using the values of $\beta$ which are obtained
by the variational method in the relativistic quark model,
we obtain the decay constants
$f_P$ and $f_V$ of the heavy mesons,
and the corresponding ratios $f_V/f_P$.
The results are compared with other calculations and
existing experimental results.
\\

\vfill 

\noindent
$1$: e-mail: dshwang@phy.sejong.ac.kr\\
$2$: e-mail: gkim@phy.sejong.ac.kr
\thispagestyle{empty} 
\pagebreak 

\baselineskip 22pt

We study the decay constants of heavy pseudoscalar and vector mesons
in the relativistic mock meson model of Godfrey, Isgur, and Capstick
\cite{gi,godf,cg}, in which
the heavy meson state composed of a
light quark $q$ and a heavy antiquark $\bar{Q}$
is represented as
\begin{equation}
|M({\bf{K}})>=\int d^3p\, \Phi ({\bf{p}})\,
{\chi}_{s{\bar{s}}}\,
{\phi}_{c{\bar{c}}}\,
|q({m_q\over m}{\bf{K}}+{\bf{p}},s)\,
{\bar{Q}}({m_{\bar{Q}}\over m}{\bf{K}}-{\bf{p}},{\bar{s}})>,
\label{f1}
\end{equation}
where $\bf{K}$ is the mock meson momentum,
$m\equiv m_q+m_{\bar{Q}}$,
and $\Phi ({\bf{p}})$, ${\chi}_{s{\bar{s}}}$, and
${\phi}_{c{\bar{c}}}$ are momentum, spin, and color wave
functions respectively.
We take the momentum wave function $\Phi ({\bf{p}})$ as a
Gaussian wave function
\begin{equation}
\Phi ({\bf{p}})={1\over ( \sqrt{\pi} \beta )^{3/2}}
e^{-{\bf{p}}^2/2{\beta}^2}.
\label{f2}
\end{equation}
The decay constant of the pseudoscalar and vector mesons, $f_P$ and
$f_V$ respectively, are defined by
\begin{equation}
<0|\, {\bar{Q}}{\gamma}^{\mu}{\gamma}_5q\, |M_P({\bf{K}})>=f_PK^{\mu},\ \ \
<0|\, {\bar{Q}}{\gamma}^{\mu}q\, |M_V({\bf{K}},\,\varepsilon )>
=f_Vm_V{\varepsilon}^{\mu}.
\label{f3}
\end{equation}
The meson state in (\ref{f1}) is written explicitly in the meson
rest frame (where ${\bf p}_q=-{\bf p}_{\bar{Q}}$) as
\begin{eqnarray}
|M_P({\bf 0})>&=&{\sqrt{2m_P}}\, \int
{d^3p_q\over
(2\pi )^{3/2} {\sqrt{2E_q\, 2E_{\bar{Q}}}}}
\,\, \Phi ({\bf p}_q)\,\,
{1\over {\sqrt{N_c}}}
\nonumber\\
& &\times {1\over {\sqrt{2}}}\,
[a_{\uparrow}^{\dagger}({\bf p}_q,c)
b_{\downarrow}^{\dagger}({\bf p}_{\bar{Q}},{\bar{c}})
-a_{\downarrow}^{\dagger}({\bf p}_q,c)
b_{\uparrow}^{\dagger}({\bf p}_{\bar{Q}},{\bar{c}})]
\, |0>,
\label{f4}
\end{eqnarray}
where the arrow indicates a state with spin up (down) along a
fixed axis and $c$ is the colour index which is summed.
Whereas we wrote the pseudoscalar meson state in (\ref{f4}), we can also write
the vector meson state in the same way with the spin combinations for
the vector states, which are given by
$(\uparrow \uparrow )$,
$1/{\sqrt{2}}\, (\uparrow \downarrow +\downarrow \uparrow )$ and
$(\downarrow \downarrow )$.
In (\ref{f4}) we adopted the normalization of the creation and annihilation
operators given by
$\{ a({\bf p},s),a^{\dagger}({\bf p}',s')\} =(2\pi )^3\, 2E\, {\delta}_{ss'}
{\delta}^3({\bf p}-{\bf p}')$, and then the meson state in (\ref{f4}) is
normalized by $<M_P({\bf 0})|M_P({\bf 0})>=2m_P\, {\delta}^3({\bf 0})$,
and also in the same way for the vector meson states.

Since we are concerned with the matrix elements in the left hand sides of
(\ref{f3}) with the meson states in (\ref{f4}), it is convenient to
represent the meson states by
\begin{equation}
{\Psi}_P\equiv - <0|\, q{\bar{Q}}\, |M_P({\bf 0})>,\ \ \
{\Psi}_V\equiv - <0|\, q{\bar{Q}}\, |M_V({\bf 0})>,
\label{f5}
\end{equation}
with which the formulas in (\ref{f3}) are written as
\begin{equation}
Tr(\, {\gamma}^{0}{\gamma}_5\, {\Psi}_P\, )=f_Pm_P,\ \ \
Tr(\, {\gamma}^{\mu}{\Psi}_V\, )=f_Vm_V{\varepsilon}^{\mu}.
\label{f6}
\end{equation}
If both two quarks inside the meson are static,
the spinor combinations of $u({\bf 0}){\bar{v}}({\bf 0})$
for the pseudoscalar and vector meson states
are given respectively as \cite{falk,neubert1}
\begin{equation}
P({\bf 0},{\bf 0})=
-{1\over {\sqrt{2}}}{1+{\gamma}^0\over 2}{\gamma}^5,\ \ \
V({\bf 0},{\bf 0},\varepsilon )=
{1\over {\sqrt{2}}}{1+{\gamma}^0\over 2}\not{\varepsilon},
\label{f7}
\end{equation}
where the polarization vectors of the vector meson are given by
${\varepsilon}^{\mu}_{\pm}=(1/{\sqrt{2}})(0,1,\pm i,0)$ and
${\varepsilon}^{\mu}_3=(0,0,0,1)$.
However, since the quarks inside the mock meson represented by
(\ref{f4}) are not static, we boost the spinors by using the 
formulas
\begin{equation}
u^{({\alpha})}(k)=
{\not{k} +m\over {\sqrt{2m(m+E)}}}u^{({\alpha})}(m,{\bf 0}),\ \ \
{\bar{v}}^{({\alpha})}(k)=
{\bar{v}}^{({\alpha})}(m,{\bf 0}){-\not{k} +m\over {\sqrt{2m(m+E)}}}.
\label{f8}
\end{equation}
Then, through this procedure we obtain
${\Psi}_P$ and ${\Psi}_V$ in (\ref{f5}) as 
\begin{equation}
{\Psi}_I=
{\sqrt{2m_I}}\int {d^3p_q\over (2\pi )^{3/2}}\, \Phi ({\bf{p}}_q)
{\sqrt{N_c}\over \sqrt{2E_q\, 2E_{\bar{Q}}}}
\, {{\not{p}}_q+m_q\over {\sqrt{2m_q(m_q+E_q)}}}
S_I
{-{\not{p}}_{\bar{Q}}+m_{\bar{Q}}\over
{\sqrt{2m_{\bar{Q}}(m_{\bar{Q}}+E_{\bar{Q}})}}},
\label{f9}
\end{equation}
where $I=P$ or $V$, and $S_P$ and $S_V$ are respectively
$P({\bf 0},{\bf 0})$ and
$V({\bf 0},{\bf 0},\varepsilon )$ in (\ref{f7}).
By incorporating (\ref{f9}) into (\ref{f6}),
we obtain the following formulas for the
decay constants of pseudoscalar and vector mesons in the
relativistic mock meson model:
\begin{equation}
f_I=
{2{\sqrt{3}}\over {\sqrt{m_I}}}\int {d^3p\over (2\pi )^{3/2}}\,
\Phi ({\bf{p}})\Big(
{E_q+m_q\over 2E_q}\, {E_{\bar{Q}}+m_{\bar{Q}}\over 2E_{\bar{Q}}}
{\Big)}^{1/2}
\,\Big( \, 1\, +\, a_I\,
{{\bf{p}}^2\over (E_q+m_q)(E_{\bar{Q}}+m_{\bar{Q}})}\, \Big) ,
\label{f10}
\end{equation}
where $I=P$ or $V$, and
\begin{equation}
a_P=-1,\ \ \ a_V=+{1\over 3}.
\label{f11}
\end{equation}
We note that the formula for $f_P$ in (\ref{f10}) and
(\ref{f11}) was already obtained by Godfrey in Ref. \cite{godf},
however we derived the formulas for both $f_P$ and $f_V$
systematically through the above procedure.

We can calculate $f_P$ and $f_V$ by using (\ref{f10}) and
(\ref{f11}). When the meson and quark masses are given,
$f_P$ and $f_V$ depend on the value of the parameter $\beta$
in (\ref{f2}).
We obtained numerically $f_P$ and $f_V$
of $B_s$, $B_d$, $D_s$, and $D_d$ mesons
as functions of
$\beta$ by using the meson \cite{rpp}
and the current quark \cite{dr} masses given by
\begin{eqnarray}
& &m_{B_s}= 5.375\, {\rm{GeV}},\
m_{B_d}= 5.279\, {\rm{GeV}},\
m_{D_s}= 1.969\, {\rm{GeV}},\
m_{D_d}= 1.869\, {\rm{GeV}},
\nonumber\\
& &m_{B_s^*}= 5.422\, {\rm{GeV}},\
m_{B_d^*}= 5.325\, {\rm{GeV}},\
m_{D_s^*}= 2.110\, {\rm{GeV}},\
m_{D_d^*}= 2.010\, {\rm{GeV}},
\nonumber\\
& &m_b= 4.397\, {\rm{GeV}},\
m_c= 1.306\, {\rm{GeV}},\
m_s= 0.199\, {\rm{GeV}},\
m_d= 0.0099\, {\rm{GeV}}.
\label{g12}
\end{eqnarray}
In Fig. 1 and 2 we present the results for the heavy mesons
having $s$ and $d$ quarks as their light quark, repectively,
and in Fig. 3 the ratios $f_V/f_P$.

In order to get the reliable results for the $B$ and $D$ meson
decay constants in our calculation, it is necessary to know
the accurate values of the parameter $\beta$.
Capstick and Godfrey used the values of $\beta$ obtained from
various methods \cite{gi,godf,cg,isgw} in this context.
In this Letter we calculate the values of $\beta$
by applying the variational
method to the relativistic hamiltonian \cite{hkn,hk}
\begin{equation}
H={\sqrt{{\bf p}^2+{m_q}^2}}+{\sqrt{{\bf p}^2+{m_{\bar{Q}}}^2}}+V(r),
\label{g13}
\end{equation}
where ${\bf r}$ and ${\bf p}$ are the relative coordinate and its
conjugate momentum.
The hamiltonian in (\ref{g13}) represents the energy of the meson in
the meson rest frame, since in this reference frame the momentum
of each quark is the same in magnitude as
that of the conjugate momentum of the relative coordinate.
In this variational calculation for $\beta$, we took 
the potential $V(r)$ in (\ref{g13})
from six different potential models,
and the results of $\beta$ have been presented in
Ref. \cite{hk}.
With these values of $\beta$ and the meson and quark masses
given in (\ref{g12}), except for the heavy quark masses ($m_b$
and $m_c$) for which we used the values given in each potential
model as the fitted parameter values,
we calculated the
decay constants by using the formulas in
(\ref{f10}) and (\ref{f11}).
The results we obtained are as follows in MeV unit:
\begin{eqnarray}
f_{B_s}= 204 \pm 7,\,\,
f_{B_d}= 173 \pm 7,\,\,
f_{D_s}= 233 \pm 12,\,\,
f_{D_d}= 191 \pm 12;& &
\nonumber\\
f_{B_s^*}= 225 \pm 9,\,\,
f_{B_d^*}= 194 \pm 8,\,\,
f_{D_s^*}= 298 \pm 11,\,\,
f_{D_d^*}= 262 \pm 10.& &
\label{g14}
\end{eqnarray}
We also present the detailed results in Table 1,
from which we get the ratios of the pseudoscalar and
vector meson decay constants:
\begin{equation}
{f_{B_s^*}\over f_{B_s}}=1.11\pm 0.00,\,\,
{f_{B_d^*}\over f_{B_d}}=1.12\pm 0.01,\,\,
{f_{D_s^*}\over f_{D_s}}=1.28\pm 0.03,\,\,
{f_{D_d^*}\over f_{D_d}}=1.37\pm 0.04.
\label{g15}
\end{equation}
We see that the ratios for $D$ mesons are bigger than those
for $B$ mesons, which can be understood by the fact that the
second term in the last factor of (\ref{f10}) contributes
more in the case of $D$ mesons.
(\ref{f10}) also gives the inequality
\begin{equation}
{\sqrt{m_V}}f_V\ge {\sqrt{m_P}}f_P,
\label{g16}
\end{equation}
in which the equality holds
in the static limit, that is,
when two quarks inside the meson are static.
Isgur and Wise obtained the similar relation,
$m_Vf_V=m_Pf_P$ (which is written in our convention of $f_V$),
in the static limit through
the heavy quark effective theory \cite{iw}.
We compare our results in (\ref{g15}) with other calculations
in Table 2:
Neubert's calculation by the heavy quark effective theory
\cite{neubert1},
and the lattice results of the ELC group \cite{ELC}.

We also compare our results by summarizing the results
of the pseudoscalar meson decay constants from
various different calculations in Table 3.
The second row is the results of the original calculations
of Capstick and Godfrey \cite{cg}, which were obtained by the
same method as ours, but with the different values of
the parameter $\beta$.
In this Letter, by obtaining the formulas for $f_P$ and $f_V$
in (\ref{f10}) and (\ref{f11}) through the formalism of
(\ref{f6}) and (\ref{f9}), we could calculate
both pseudoscalar and vector meson decay constants.
Dominguez did the calculation using the QCD sum rules
\cite{doming}.
The fourth and fifth rows are from the lattice calculations
of the UKQCD \cite{UKQCD} and the BLS \cite{BLS} groups.
In the last two rows
we wrote the experimental results of the WA75
\cite{wa75} and the CLEO \cite{cleo94} groups.
We find that our result of $f_{D_s}$ agrees very well with
the WA75 result.
Finally we mention that we get the value of the double ratio
$(f_{B_s}/f_{B_d})\, /\, (f_{D_s}/f_{D_d})$
as $0.967$ from our results in Table 3,
which is exactly the same as the value
that Grinstein obtained by the heavy quark effective theory
\cite{grin}.
\\
 
\noindent
{\em Acknowledgements} \\
\indent
This work was supported
in part by the Basic Science Research Institute Program,
Ministry of Education, Project No. BSRI-94-2414,
and in part by Daeyang Foundation at Sejong University.\\
 
\pagebreak

\pagebreak

\begin{table}[h]
\vspace*{1.2cm}
\hspace*{-1.2cm}
%\begin{center}
\begin{tabular}{|c|c|c|c|c|c|c|c|c|}   \hline
Model&$f_{B_s}$&$f_{B_s^*}$
&$f_{B_d}$&$f_{B_d^*}$
&$f_{D_s}$&$f_{D_s^*}$
&$f_{D_d}$&$f_{D_d^*}$\\   \hline
A ( Eich. )\cite{eich}    &198&218&170&187&242&300&200&264\\
B ( Hagi. )\cite{hagi}    &190&209&160&180&210&276&170&243\\
C (Power 1)\cite{martin}  &208&229&177&199&249&313&207&276\\
D (Power 2)\cite{ros93}   &213&237&181&204&234&301&190&262\\
E ( Log.  )\cite{qr}      &206&229&175&198&229&296&187&259\\
F ( Rich. )\cite{richard} &206&228&175&197&231&300&190&265\\
\hline
(Average)  &204$\pm 7$&225$\pm 9$&173$\pm 7$&194$\pm 8$
&233$\pm 12$&298$\pm 11$&191$\pm 12$&262$\pm 10$\\
\hline
\end{tabular}
%\end{center}
\caption{The pseudoscalar and vector meson decay constants
(MeV) in six different potential models.}
\end{table}

\begin{table}[h]
\vspace*{1.2cm}
\begin{center}
\begin{tabular}{|c|c|c|c|c|}   \hline
    &$f_{B_s^*}/f_{B_s}$&$f_{B_d^*}/f_{B_d}$
&$f_{D_s^*}/f_{D_s}$&$f_{D_d^*}/f_{D_d}$\\  \hline
This Work
&1.11$\pm 0.00$&1.12$\pm 0.01$&1.28$\pm 0.03$&1.37$\pm 0.04$\\
Neubert \cite{neubert1}&---&1.07$\pm 0.02$&---&1.35$\pm 0.05$\\
ELC \cite{ELC}&---&1.12$\pm 0.05$&---&1.30$\pm 0.06$\\
\hline
\end{tabular}
\end{center}
\caption{The ratios of the decay constants of vector and
pseudoscalar mesons.}
\end{table}

\begin{table}[h]
\vspace*{1.2cm}
\hspace*{-2.5cm}
%\begin{center}
\begin{tabular}{|c|c|c|c|c|c|c|}   \hline
     &$f_{B_s}$&$f_{B_d}$&$f_{B_s}/f_{B_d}$
&$f_{D_s}$&$f_{D_d}$&$f_{D_s}/f_{D_d}$\\  \hline
This Work&204$\pm 7$&173$\pm 7$&1.18$\pm .01$
&233$\pm 12$&191$\pm 12$&1.22$\pm .01$\\
 
Cap. Godf. \cite{cg} &210$\pm 20$&155$\pm 15$&1.35$\pm .18$
&290$\pm 20$&240$\pm 20$&1.21$\pm .13$\\

Doming. \cite{doming}&193$\pm 28$&158$\pm 25$&1.22$\pm .02$
&222$\pm 48$&187$\pm 48$&1.21$\pm .06$\\

%(HQET)\cite{}& & & & & & \\
% 
UKQCD\cite{UKQCD}&194${\, }^{+6+62}_{-5-9}$
&160${\, }^{+6+53}_{-6-19}$&1.22${\, }^{+.04}_{-.03}$
&212${\, }^{+4+46}_{-4-7}$&185${\, }^{+4+42}_{-3-7}$
&1.18$\pm .02$\\
 
BLS  \cite{BLS} &207$\pm 9\pm 40$&187$\pm 10\pm 37$
&1.11$\pm .02\pm .05$
&230$\pm 7\pm 35$&208$\pm 9\pm 37$&1.11$\pm .02\pm .05$\\
\hline

WA75 \cite{wa75}&---&---&---
&232$\pm 45\pm 52$&---&---\\

CLEO \cite{cleo94} &---&---&---
&344$\pm 37\pm 67$&---&---\\  \hline
\end{tabular}
%\end{center}
\caption{The values (MeV) and ratios of the decay constants from
different calculations and experimental results.}
\end{table}

\pagebreak

\vspace*{3.5cm}
\vspace*{14.5cm}
\noindent
Fig. 1.
$f_{B_s^*}$,
$f_{B_s}$,
$f_{D_s^*}$, and
$f_{D_s}$
as functions of the parameter $\beta$.

\pagebreak

\vspace*{3.5cm}
\vspace*{14.5cm}
\noindent
Fig. 2.
$f_{B_d^*}$,
$f_{B_d}$,
$f_{D_d^*}$, and
$f_{D_d}$
as functions of the parameter $\beta$.

\pagebreak

\vspace*{3.5cm}
\vspace*{14.5cm}
\noindent
Fig. 3.
$f_{B_s^*}/f_{B_s}$,
$f_{B_d^*}/f_{B_d}$,
$f_{D_s^*}/f_{D_s}$, and
$f_{D_d^*}/f_{D_d}$
as functions of the parameter $\beta$.

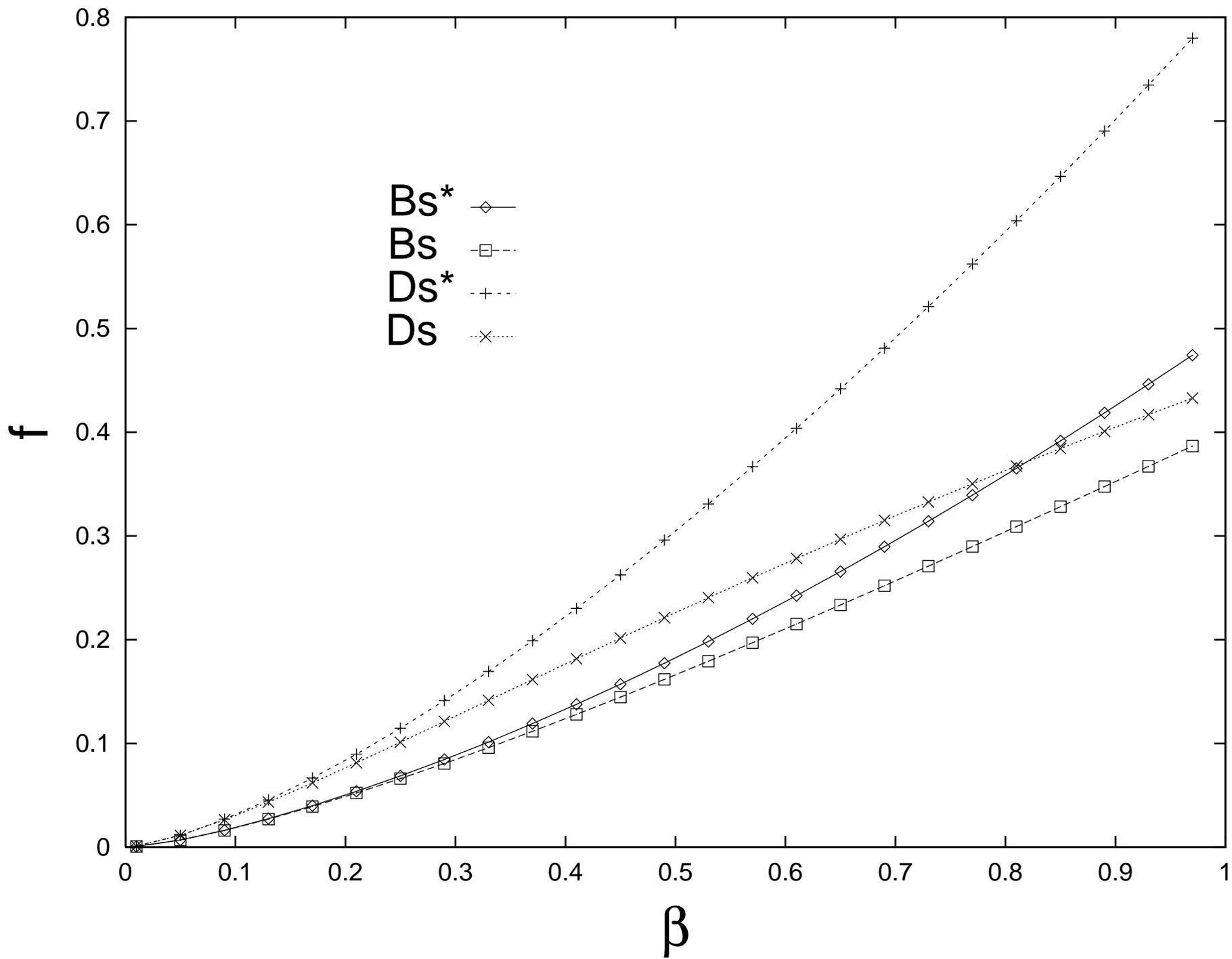
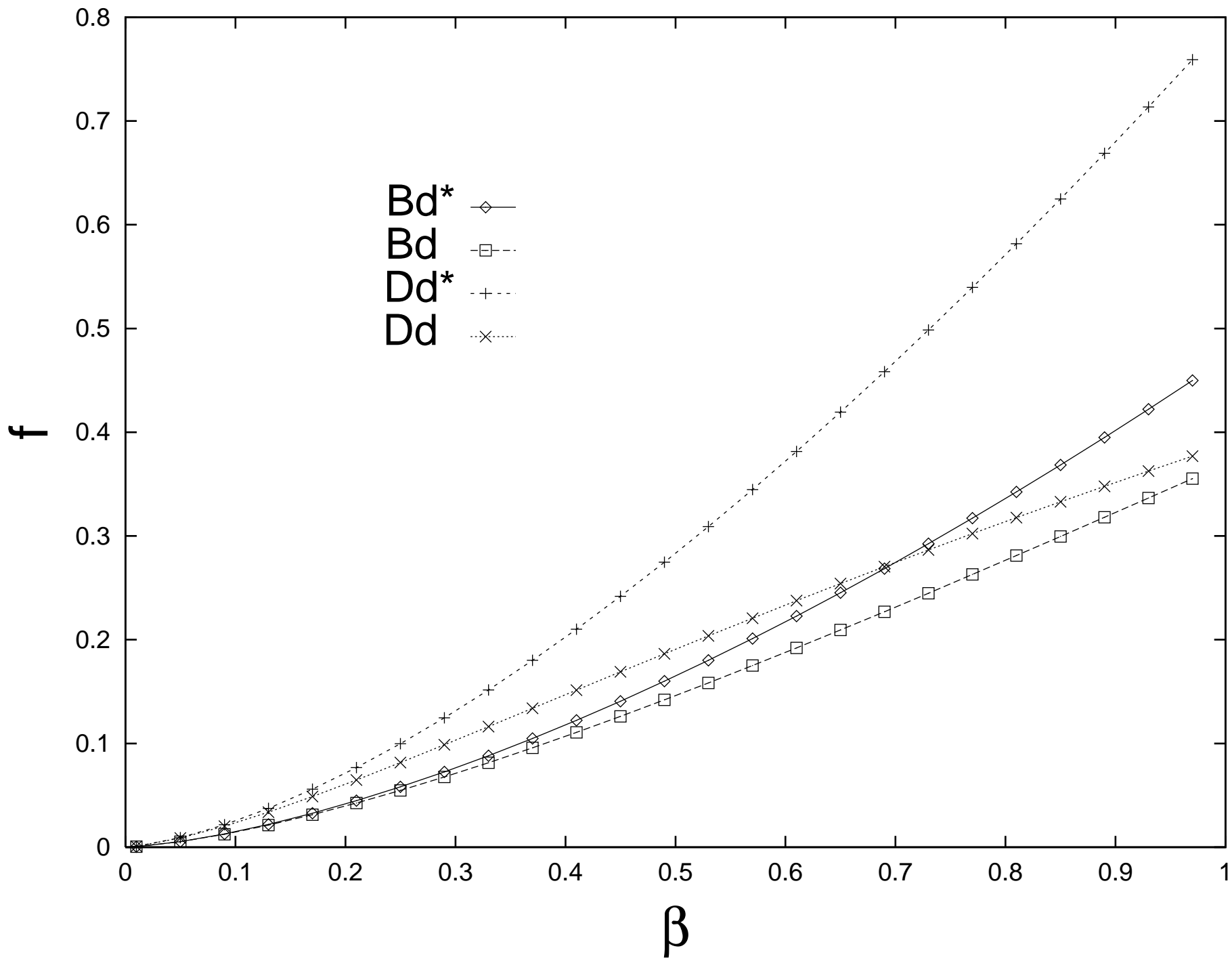
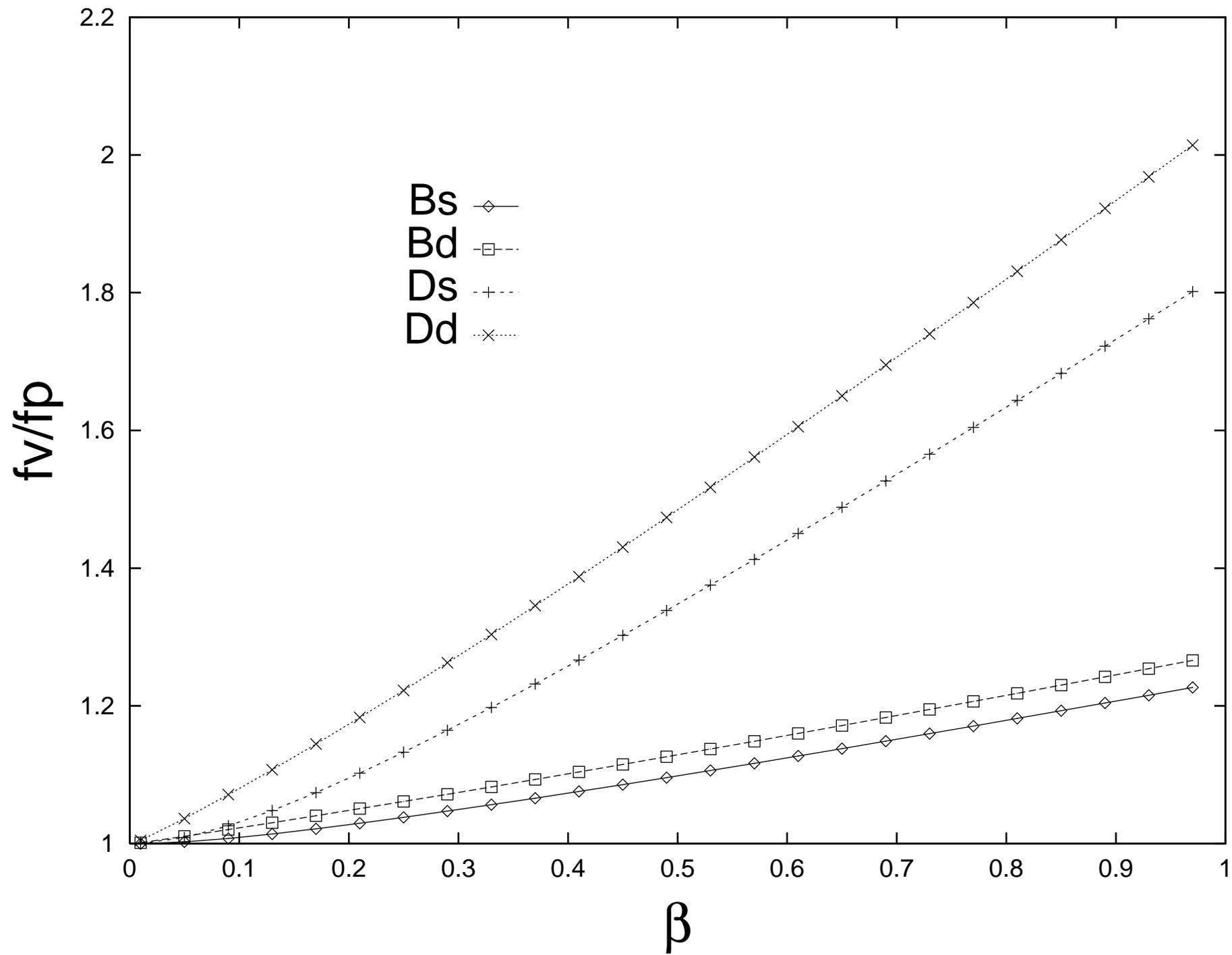
\end{document}